\documentclass[12pt]{article}

\newcommand{\x}{arXiv:}
\newcommand{\m}{\mathrm}
\newcommand{\be}{\begin{equation}}
\newcommand{\ee}{\end{equation}}
\newcommand{\ba}{\begin{eqnarray}}
\newcommand{\ea}{\end{eqnarray}}

\usepackage{graphicx}
\usepackage{amssymb}
\usepackage{amsmath}
\usepackage[T1]{fontenc} 
\usepackage[ansinew]{inputenc} 
\usepackage[nosort]{cite}
\newcommand{\inbar}{\vrule height1.57ex width.4pt depth0pt}
\newcommand{\SW}{\relax{\hbox{$\ \inbar\kern-.285em{\rm S}$}}}

\begin{document}
\thispagestyle{empty}
\begin{center}

\null \vskip-1truecm \vskip2truecm

{\Large{\bf \textsf{A Rotation/Magnetism Analogy for the Quark-Gluon Plasma}}}

{\Large{\bf \textsf{}}}

{\large{\bf \textsf{}}}

{\large{\bf \textsf{}}}

\vskip1truecm

{\large \textsf{Brett McInnes
}}

\vskip0.1truecm

\textsf{\\ National
  University of Singapore}
  \vskip1.2truecm
\textsf{email: matmcinn@nus.edu.sg}\\

\end{center}
\vskip1truecm \centerline{\textsf{ABSTRACT}} \baselineskip=15pt

\medskip

In peripheral heavy ion collisions, the Quark-Gluon Plasma that may be formed often has a large angular momentum per unit energy. This angular momentum may take the form of (local) rotation. In many physical systems, rotation can have effects analogous to those produced by a magnetic field; thus, there is a risk that the effects of local rotation in the QGP might be mistaken for those of the large genuine magnetic fields which are also known to arise in these systems. Here we use the gauge-gravity duality to investigate this, and we find indeed that, with realistic parameter values, local rotation has effects on the QGP (at high values of the baryonic chemical potential) which are not only of the same kind as those produced by magnetic fields, but which can in fact be substantially larger. Furthermore, the combined effect of rotation and magnetism is to change the shape of the main quark matter phase transition line in an interesting way, reducing the magnitude of its curvature; again, local rotation contributes to this phenomenon at least as strongly as magnetism.

\newpage
\addtocounter{section}{1}
\section* {\large{\textsf{1. Rotation/Magnetism and the Quark-Gluon Plasma}}}

It has often been observed that, in many physical systems, local rotation (or vorticity) plays a role analogous to that of a magnetic field: to take but one of many examples, this ``rotation/magnetism analogy'' is important in the study of the quantum Hall effect \cite{kn:viefers}.

As is now well known \cite{kn:skokov,kn:tuchin,kn:magnet,kn:review}, huge magnetic fields can be present in the quark-gluon plasma (QGP) produced by peripheral heavy-ion collisions \cite{kn:andronicover,kn:ALICE,kn:braun,kn:armesto,kn:vuor}, and these can give rise to a number of remarkable effects. In particular, various computations suggest that \emph{strong magnetic fields tend to lower the temperature} at which various phenomena are otherwise expected to occur. For example, lattice computations \cite{kn:bali} indicate the existence of a very remarkable ``inverse magnetic catalysis'' effect, in which the presence of a strong magnetic field lowers the temperature of the chiral transition (and, presumably ---$\,$ but see \cite{kn:dudal} ---$\,$ also the pseudo-critical temperature): see \cite{kn:naylor,kn:fraga}.

On the other hand, it has recently become clear that local rotation might \emph{also} be important in these systems, and this might manifest itself in the form of such phenomena as the ``chiral vortical effect'': see \cite{kn:volosh} for a review. Now the large magnetic fields in the QGP mentioned above are in fact closely associated with very large angular momentum densities \cite{kn:liang,kn:bec,kn:huang,kn:KelvinHelm,kn:viscous,kn:csernairecent1,kn:csernairecent2,kn:nagy,kn:nacs}; see \cite{kn:yin,kn:deng} for recent in-depth analyses. The angular momentum arises in the same way as the magnetic field, and the corresponding vectors are (to a good approximation) parallel \cite{kn:deng} (that is, perpendicular to the reaction plane).

This prompts the question: could the rotation/magnetism analogy be valid for the QGP? Might, for example, local rotation directly affect temperatures, just as magnetism apparently does? If this is so, then \emph{ignoring the effects of local rotation could lead to serious errors} in estimating the effects of the magnetic field on the behaviour of the plasma.

For example, suppose that one has a calculation, for example lattice-based, of the likely location of the quark matter critical endpoint (see for example \cite{kn:mohanty,kn:satz,kn:race,kn:karsch,kn:magdy,kn:incera}) in the quark matter phase diagram. For the QGP produced in peripheral collisions, it is thought that the corresponding magnetic field lowers the temperature and the baryonic chemical potential, $\mu_B$, at that point (with fixed values of the other parameters), relative to the values expected in the absence of a magnetic field ---$\,$ that is, in a central collision. (See \cite{kn:magdy,kn:incera} and references therein for recent discussions of this.) But if there is a rotation/magnetism analogy, the local rotation generated by a peripheral collision could have an independent effect which might significantly strengthen (\emph{or even weaken}) this important phenomenon. (Note that rotation itself may play a useful role \cite{kn:simple} in locating the critical endpoint, underlining the importance of understanding its effects.)

To take another, related example: the quark matter critical endpoint, if it exists, is associated with a \emph{phase transition line}, and investigating this line (see Figure 1) is a prime objective of the beam energy scan experiments such as RHIC, GSI-FAIR and NICA, among others \cite{kn:STAR,kn:shine,kn:nica,kn:fair,kn:BEAM,kn:luo}.
\begin{figure}[!h]
\centering
\includegraphics[width=0.95\textwidth]{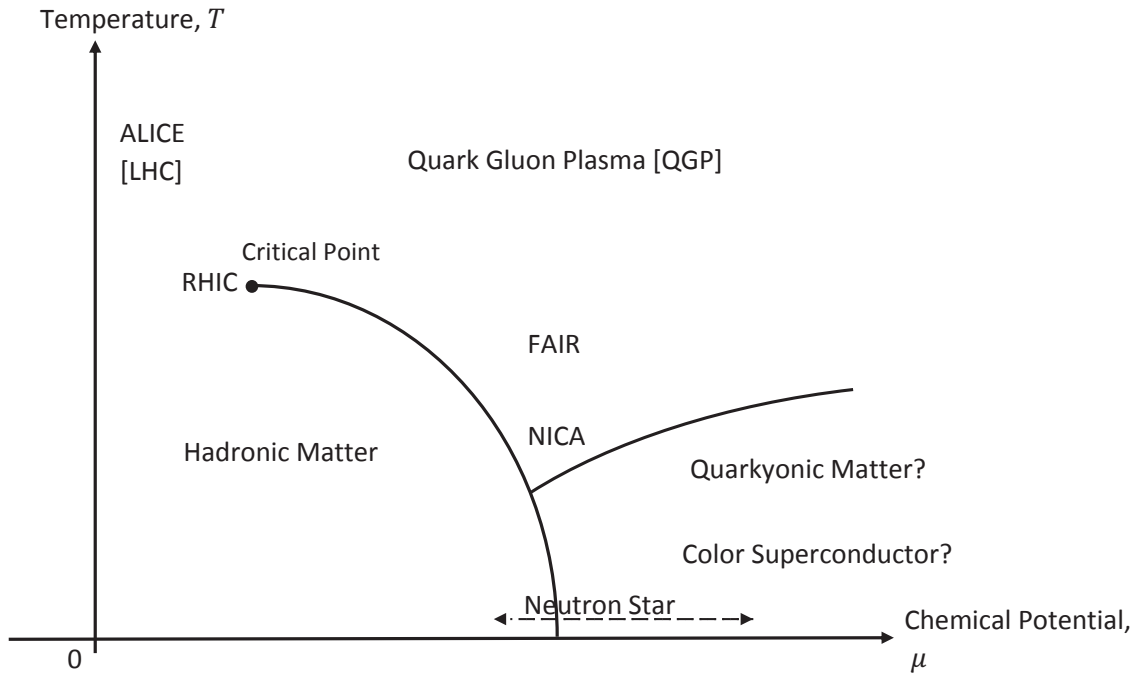}
\caption{Quark Matter Phase Diagram. }
\end{figure}
The line\footnote{There may of course be other phase transition lines, as shown in Figure 1, but we will not consider them here.} bends down into the region of lower temperature and higher $\mu_B$, that is, into the region of higher net particle density. From a theoretical point of view, one hopes ultimately to compute not just the location of the critical endpoint but also the \emph{curvature} of this line. Now, strong magnetic fields might affect not just the location but also the shape of the phase line. In other words, there could be a non-trivial interaction between increased net particle density and magnetism. However, if there is indeed a local rotation/magnetism analogy, there will be a corresponding interaction between the rotational angular momentum and the net particle density\footnote{This could be related to the recent suggestion of a possible ``rotation/density'' analogy \cite{kn:fuku}.}. Again, if this is indeed so, then ignoring local rotation could lead to erroneous predictions regarding this entire region of the phase diagram, for quark matter produced in peripheral collisions.

In short, if a local rotation/magnetism analogy exists, it must be taken into account in theoretical studies related to the high-$\mu_B$ experiments currently under way or projected: it is quite possible that analyses which give good results for central collisions may otherwise fail when extended to the peripheral case.

The strongly coupled QGP is, however, not well-understood when $\mu_B$ is comparable to or larger than the temperature, and it is not clear that the local rotation/magnetism analogy works in any straightforward way here. Theoretical investigations of the effects of local rotation can be pursued using lattice methods \cite{kn:rotation}, but, at large values of $\mu_B$, one encounters the usual ``sign problem''. We wish to argue that the gauge-gravity (``holographic'') correspondence \cite{kn:youngman,kn:gubser,kn:janik}, in which a given problem regarding the plasma is related to an equivalent problem in a dual gravitational system, may be useful here.

In the simplest cases, the gauge-gravity duality postulates a duality of some ``plasma-like'' boundary theory with a suitably chosen asymptotically AdS black hole in the bulk\footnote{The bulk here is four-dimensional, since the dual physics is effectively three-dimensional: the rotational motion is studied by confining attention to the reaction plane (with the standard $(x, z)$ coordinates), which is strictly perpendicular to the magnetic field and the angular momentum vectors (both parallel to the usual $y$ axis.)}. In the application at hand, this black hole must rotate, to reflect the rotation at infinity, and it must be charged (both electrically and magnetically), to reflect non-zero values of the magnetic field and of the baryonic chemical potential in the dual field theory. A holographic study of the local rotation/magnetism analogy, if it is valid in this application, would therefore involve the study of one of the (several) asymptotically AdS black holes which are counterparts of the (dyonic) asymptotically flat Kerr-Newman \cite{kn:visser} geometry\footnote{Asymptotically AdS black holes endowed with angular momentum have been studied with a view to holographic applications in \cite{kn:sonner,kn:schalm,kn:75,kn:shear,kn:79,kn:84}. The holography of inverse magnetic catalysis was considered in \cite{kn:mamo,kn:rouge,kn:86}.}.

A holographic treatment of angular momentum is possible because of two effects: first, as is well known, a black hole endowed with angular momentum induces \emph{frame-dragging} in the ambient spacetime; and, second, because this frame-dragging persists all the way to infinity in the case of asymptotically AdS black holes. Unlike their asymptotically flat counterparts, these are not uniquely specified by their metric parameters: the geometry of the event horizon need not be that of a sphere \cite{kn:lemmo}. Broadly speaking, there are two well-behaved\footnote{``Well-behaved'' in the sense of being dual to a well-defined (stable) boundary theory.} classes: first, black holes with event horizons which (prior to the introduction of angular momentum) have the local geometry of a sphere, and, second, black holes with event horizons which (prior to the introduction of angular momentum) have the local geometry of a flat plane.

It turns out, very remarkably, that this classification precisely reflects the two basic ways \cite{kn:KelvinHelm} in which angular momentum is manifested in the QGP in the aftermath of a heavy-ion collision: as local rotation (vorticity) \cite{kn:viscous,kn:csernairecent1,kn:csernairecent2,kn:nagy,kn:nacs,kn:yin}, or as a shearing motion \cite{kn:liang,kn:bec,kn:huang,kn:deng}. (The two are not mutually exclusive, and in fact in a real plasma the two forms would co-exist, but for clarity we treat them separately.) Here we are concerned with local rotation or vorticity, and one finds that this corresponds to the case of asymptotically AdS black holes with topologically \emph{spherical} event horizons: that is, to the simplest extension of the asymptotically flat dyonic Kerr-Newman metric to the AdS context. Even this ``simplest'' generalization can give rise to surprises, however.

We are interested in studying the temperature of the plasma and its baryonic chemical potential. The quantity which is holographically dual to the temperature of the QGP is of course the Hawking temperature of the bulk black hole, which depends both on its electromagnetic charges \emph{and} on its angular momentum. As we will see, the dual of the baryonic chemical potential is also a certain combination of the charges \emph{and} the angular momentum parameter. In this very general sense, holography indicates that there must indeed be some kind of analogy between local rotation and a magnetic field in the dual plasma. Notice that an observation which is immediate on one side of the duality is far from it on the other: this is the point of holography.

These considerations allow us to formulate the analogy in a concrete way; so we can begin to answer some basic questions. Does local rotation always change the temperature and chemical potential \emph{in the same direction} as does a magnetic field ---$\,$ that is, does it always reinforce the magnetic effect or sometimes tend to counteract it\footnote{In fact, in the case of \emph{shearing} angular momentum, it was found in \cite{kn:86} that unexpected behaviour does occur in some cases: it can happen that increasing the shearing angular momentum has the opposite effect to that of a large magnetic field (though never enough to counteract the latter entirely ---$\,$ in this sense, shearing angular momentum has a  ``weaker'' effect than magnetism).}? Which of the two effects is dominant? Do local rotation and magnetism modify the shape of the phase transition line, and, if so, how do the two effects compare?

In fact, the problem of understanding the behaviour of the temperature and chemical potential corresponding to a dyonic, topologically spherical AdS black hole is not straightforward: for two reasons. First, unlike the parameters of an asymptotically flat black hole, the geometric parameters of a rotating AdS black hole are not related in any simple way to its physical properties. For example, the mass parameter $M$ is \emph{not} the physical mass of the black hole (see equations (\ref{C}) below). Second, holography does not relate the boundary parameters to the physical parameters of the black hole in as simple a way as in the non-rotating, non-magnetic case: for example, the baryonic chemical potential of the boundary theory is \emph{not} proportional to the electric charge on the black hole here (see equation (\ref{L}) below), as it is in the familiar case of the dyonic AdS-Reissner-Nordstr\"om black hole.

In short, the answers to our questions are unclear: they can only be settled by means of a detailed investigation, which we propose to carry out here. We stress that the holographic model of these extremely complex systems may well be severely over-simplified, so that accurate numerical predictions are not to be expected here. Our emphasis is on more qualitative questions. The answers to these questions may guide more quantitative investigations of these basic properties of the QGP.

In order to proceed, we need a detailed description of gauge-gravity duality in this case, including a ``dictionary'' that converts all of the relevant boundary parameters to quantities describing an asymptotically AdS black hole. To this we now turn.

\addtocounter{section}{1}
\section* {\large{\textsf{2. The Holographic Dictionary for the Topologically Spherical Dyonic AdS-Kerr-Newman black hole.}}}
The dyonic (that is, both electrically and magnetically charged) four-dimensional AdS-Kerr-Newman black hole with topologically spherical event horizon has a metric of the form, in Boyer-Lindquist-like coordinates \cite{kn:cognola},
\begin{flalign}\label{A}
g(\m{AdSdyKN^{+1}_{4})} = &- {\Delta_r \over \rho^2}\Bigg[\,\m{d}t \; - \; {a \over \Xi}\m{sin}^2\theta \,\m{d}\phi\Bigg]^2\;+\;{\rho^2 \over \Delta_r}\m{d}r^2\;+\;{\rho^2 \over \Delta_{\theta}}\m{d}\theta^2 \\ \notag \,\,\,\,&+\;{\m{sin}^2\theta \,\Delta_{\theta} \over \rho^2}\Bigg[a\,\m{d}t \; - \;{r^2\,+\,a^2 \over \Xi}\,\m{d}\phi\Bigg]^2,
\end{flalign}
where the ``dy'' denotes ``dyonic'', where the ``$+1$'' reminds us of the spherical topology of the event horizon, and where
\begin{eqnarray}\label{eq:B}
\rho^2& = & r^2\;+\;a^2\m{cos}^2\theta, \nonumber\\
\Delta_r & = & (r^2+a^2)\Big(1 + {r^2\over L^2}\Big) - 2Mr + {Q^2 + P^2\over 4\pi},\nonumber\\
\Delta_{\theta}& = & 1 - {a^2\over L^2} \, \m{cos}^2\theta, \nonumber\\
\Xi & = & 1 - {a^2\over L^2}.
\end{eqnarray}
Here $- 1/L^2$ is the asymptotic curvature, and $a$ is the ``specific angular momentum'' (angular momentum per unit physical mass). Note that the quantity $\Xi$ \emph{must} be present in order to ensure regularity of the geometry: see below. The geometric parameters $M, Q$, and $P$ are related to the physical mass $E$, electric charge $q$, and magnetic charge $p$, by \cite{kn:gibperry}\cite{kn:hajian}
\begin{equation}\label{C}
E\;=\;M/\Xi^2, \;\;\;\;\;q\;=\;Q/\Xi,\;\;\;\;\;p\;=P/\Xi.
\end{equation}
We begin to see the principal point: for example, the parameter $M$ in the metric is related to the physical mass through a formula which \emph{also} involves the specific angular momentum \cite{kn:mcong}, through the quantity $\Xi$. More generally, the ubiquity of $\Xi$ throughout the formulae describing this system implies that the physical black hole parameters vary in a complicated way as the specific angular momentum is varied\footnote{We should stress that the corresponding black hole describing a shearing plasma \cite{kn:shear,kn:86} is \emph{very} different to the one considered here: in particular, there is nothing analogous to $\Xi$ in that case. The results of \cite{kn:86} therefore offer no guide as to what one should expect here.}. We note in passing that the presence of the $\Xi$ factors requires that the inequality
\begin{equation}\label{CATE}
a^2/L^2 \;<\; 1
\end{equation}
must always be satisfied.

The electromagnetic potential form here is given by
\begin{equation}\label{D}
A = \left({-\,\Xi\over 4\pi \rho^2L}\,\left[Qr+aP\,\m{cos}\theta\right]+\kappa_t\right)\,\m{d}t\;+\;\left({1\over 4\pi\rho^2L}\,\left[Qar\,\m{sin}^2\theta+P\,\m{cos}\theta\left\{r^2+a^2\right\}\right]+\kappa_{\phi}\right)\,\m{d}\phi ,
\end{equation}
where $\kappa_t$ and $\kappa_{\phi}$ are constants which can be evaluated as follows. First, consider the Euclidean version of $g(\m{AdSdyKN^{+1}_{4})}$, obtained by complexifying $t, Q,$ and $a$ (but not $P$). A calculation of their lengths in this metric shows that $\partial_t$ and $\partial_{\phi}$ vanish at the points corresponding to the poles of the Euclidean ``event horizon'' (where the Euclidean version of $\Delta_r$ has its largest root), and so we must have $A^E(\partial_t) = A^E(\partial_{\phi}) = 0$ there if the Euclidean potential $A^E$ is to be regular. These equations yield the values of the Euclidean versions of $\kappa_t$ and $\kappa_{\phi}$, and continuing back to the Lorentzian section we have finally
\begin{flalign}\label{E}
A = &\left(-\,{\Xi\over 4\pi \rho^2L}\,\left[Qr+aP\,\m{cos}\theta\right]+{\Xi\left(Q\,r_h+aP\right)\over 4\pi L \left(r_h^2+a^2\right)}\right)\,\m{d}t \\ \notag \,\,\,\,&+\;\left({1\over 4\pi\rho^2L}\,\left[Qar\,\m{sin}^2\theta+P\,\m{cos}\theta\left\{r^2+a^2\right\}\right]-{P\over 4\pi L}\right)\,\m{d}\phi,
\end{flalign}
where $r_h$ is the value of the radial coordinate at the event horizon.

The conformal boundary of this spacetime has a structure which can be represented by the metric
\begin{equation}\label{F}
g(\m{AdSdyKN^{+1}_{4})}_{\infty}\;=\;-\,\m{d}t^2 \;-\;{2a\,\m{sin}^2(\theta)\,\m{d}t \m{d}\phi\over \Xi} \;+\; {L^2 \, \m{d}\theta^2 \over 1 - (a/L)^2\m{cos}^2(\theta)} \;+\; {L^2 \m{sin}^2(\theta)\m{d}\phi^2\over \Xi};
\end{equation}
with this choice, $t$ represents proper time for a distinguished observer at infinity.

A massive, zero-momentum particle in this boundary geometry has a worldline satisfying $\dot{\phi}$ = $a\,\dot{t}/L^2$ (where the dot refers to the proper time of the particle), so it is frame-dragged in the $\phi$ direction at an angular velocity relative to the distinguished observer of $\omega$ = $a/L^2$. This shows that frame-dragging does indeed persist to infinity in this case, and that (since $\omega$ is independent of position at infinity) this frame-dragging takes the form of uniform \emph{rotation} in the reaction plane. Thus, under these conditions, we have a holographic setup for the locally rotating QGP.

Unfortunately, the boundary theory is evidently defined on a spacetime with curved spatial sections. This is obviously not realistic for the application to heavy ion collisions, and, furthermore, the curvature induces irrelevant or unphysical effects such as spurious mass gap phenomena. It is therefore essential to ensure that the curvature is negligible here. We proceed as follows.

Near to either pole, the spatial geometry induced by the metric $g(\m{AdSdyKN^{+1}_{4})}_{\infty}$ is approximately that of a round two-sphere\footnote{It is by considering this two-sphere geometry that one understands why the important quantity $\Xi$ must be present throughout this work: without it, the geometry would never be approximately flat, even for arbitrarily small regions around the poles. See \cite{kn:79} for the details.} of radius $\hat{L} = L/\sqrt{\Xi}$; therefore, this geometry will be approximately flat provided that $\hat{L}$ is sufficiently large relative to the size of the system being described, which in our case would be a sample of rotating plasma, placed so that its centre is at the pole\footnote{This ``sample'' is to be regarded as a relatively small part of the entire system produced in a peripheral collision. The vorticity is expected to vary with position \cite{kn:yin}, so our analysis should be thought of as applying locally, to each of many samples.}. In order to ensure this, we must, since $\hat{L}\,\geq L$ for all values of $a$ (with equality for $a = 0$, that is, central collisions), take $L$ to be significantly larger than the plasma sample, that is, more than about 10 femtometres. From a physical point of view, we should take it to be of the order of the largest length scale naturally associated with this system. As we will see later, this is provided by $a$, which (in this case) ranges up to about 50 femtometres: so, bearing in mind the inequality (\ref{CATE}), we will take $L \approx 100$ femtometres. This ensures that the space we are dealing with can be assumed flat under all circumstances; notice that in fact the approximation becomes a little better if $a$ is relatively large, since then $\hat{L}$ is somewhat larger than $L$. (Notice too that this value of $L$ is compatible with the conditions for holography to be valid from a string-theoretic point of view, that is, it is much larger than the string scale.)

Under these circumstances, $\varrho$ = $\hat{L}\theta$ and $\phi$ define plane polar coordinates in the flat space tangential to the pole\footnote{The presence of $\hat{L}$ here can be understood physically in terms of the well-known problem of the rotating disc \cite{kn:rizzi}: the rotation causes the circumference of the disc to increase, according to an observer on the disc, by the usual Lorentz factor. In this case, that factor is $\Xi^{-1/2} = 1/\sqrt{1-(a/L)^2}$, so we can think of $a/L$ as the maximal rotational velocity of the system; this also explains the inequality (\ref{CATE}), above. Our choice for $L$ represents a maximal velocity which is relativistic but still causal.}, and this flat space is dual to the reaction plane of the collision giving rise to the plasma. We are now in a position to construct the holographic dictionary in this case.

We are interested in the following parameters describing the plasma: its specific angular momentum, its temperature, the intensity of the magnetic field to which it is subjected, its specific entropy (entropy per unit of energy), and its baryonic chemical potential (which is related to the net particle density). We consider these in turn.

The specific angular momentum $a$ of the black hole will be interpreted as the specific angular momentum of the plasma, that is, the ratio of its angular momentum density to its energy density. This quantity varies for different collisions in a given beam, ranging from zero (in exactly central collisions) up to some maximum which can be estimated \cite{kn:bec}. We therefore have to allow for a \emph{range} of values for $a$ (from zero up to some value below $L$, as explained above).

The temperature of the plasma corresponds to the Hawking temperature of the black hole. This is given \cite{kn:cognola} by
\begin{equation}\label{FROG}
T\;=\;{r_h \Big(1\,+\,a^2/L^2\,+\,3r_h^2/L^2\,-\,{a^2\,+\,\{Q^2+P^2\}/4\pi \over r_h^2}\Big)\over 4\pi (a^2\,+\,r_h^2)},
\end{equation}
where, as before, $r = r_h$ locates the event horizon.

Next, consider the electromagnetic potential (equation (\ref{E})) at infinity: we have
\begin{flalign}\label{G}
A_{\infty} \,=\, {\Xi\left(Q\,r_h+aP\right)\over 4\pi L\left(r_h^2+a^2\right)}\,\m{d}t +\;{P\over 4\pi L}\left(\m{cos}\theta \,-\,1\right)\m{d}\phi,
\end{flalign}
with corresponding field strength
\begin{equation}\label{H}
F_{\infty}\,=\,-\,{P\over 4\pi L}\,\m{sin}\theta \,\m{d}\theta \wedge \m{d}\phi.
\end{equation}
Let $\hat{\theta}$ and $\hat{\phi}$ represent unit (relative to $g(\m{AdSdyKN^{+1}_{4})}_{\infty}$) one-forms parallel to $\m{d}\theta$ and $\m{d}\phi$. One finds that $\hat{\theta} = L\,\m{d}\theta/\sqrt{\Delta_{\theta}},\;\; \hat{\phi} = L\,\m{sin}\theta \,\,\m{d}\phi \, \sqrt{\Delta_{\theta}}/\Xi$, and so
\begin{equation}\label{I}
F_{\infty}\,=\,-\,{\Xi \,P\over 4\pi L^3}\,\hat{\theta} \,\wedge \hat{\phi}.
\end{equation}
Clearly we have here a magnetic field at infinity, perpendicular to the reaction plane and uniform within it, given by
\begin{equation}\label{J}
B\,=\,{\Xi \,P\over L^3}.
\end{equation}
This is similar to the way magnetic fields are treated in holographic condensed matter theory \cite{kn:hartkov}. As with the specific angular momentum, the value of $B$ depends on the geometry of each collision: like $a$, it varies from zero up to some maximum, so we have to consider a range of values.

Equation (\ref{J}) is the third entry in our ``dictionary''. Again we see that the relation between the black hole magnetic charge and the boundary magnetic field is not simple, for it too involves the specific angular momentum of the black hole.

Next, we turn to the specific entropy of the black hole. According to Hawking's formula, the entropy itself is given by one quarter of the area of the event horizon; this area is $4\pi (r_h^2+a^2)/\Xi$, and since the physical mass is (as above) $M/\Xi^2$, the specific entropy is
\begin{equation}\label{K}
\varsigma_S\,=\,{\pi \, \Xi\,(r_h^2 + a^2)\over M}.
\end{equation}
We interpret this holographically as the average entropy (per unit of energy) of the particles in the collisions, occurring in a given beam, that give rise to the plasma. This is a physical characteristic of the beam; in order to make a meaningful comparison, we hold it constant when we consider two situations: one in which the effects of vorticity and magnetism are ignored, and one in which they are taken into account.

Finally, the holographic version of the chemical potential is obtained by examining the electric potential as evaluated by an observer at infinity. Since $\partial_t$ is a unit vector at infinity, this means that we simply take the timelike component of $A_{\infty}$ (equation (\ref{G}) above) and so we obtain
\begin{equation}\label{L}
\mu_B \,=\,{3\Xi\left(Q\,r_h+aP\right)\over 4\pi L\left(r_h^2+a^2\right)},
\end{equation}
the factor of 3 being needed to express the relation in terms of the usual baryonic chemical potential. Notice that $\mu_B$ depends explicitly (and implicitly, through $r_h$) on \emph{both} $Q$ and $P$ ---$\,$ as well as on $a$.

Recall that the baryonic chemical potential is related to the net particle density, that is, to the extent to which particles dominate over antiparticles; see for example the discussion in \cite{kn:phobos}. However, the particle/antiparticle ratio is given not by $\mu_B$ itself (which has units of length$^{-1}$ here) but rather as a monotonically increasing function of the dimensionless ratio $\mu_B/T$. We therefore quantify ``density'' here by the quantity $\Gamma$, defined simply as
\begin{equation}\label{LONDON}
\Gamma \;=\;\mu_B/T.
\end{equation}
This quantity is directly accessible to experiment; $\mu_B$ is secondary in this sense. As with the specific entropy, this physical parameter allows us to compare the situations with and without vorticity/magnetism.

The horizon coordinate $r_h$, which occurs in many of our equations, is related to the other parameters through its definition,
\begin{equation}\label{M}
\Delta_r(r_h)\;=\;(r_h^2+a^2)\Big(1 + {r_h^2\over L^2}\Big) - 2Mr_h + {Q^2 + P^2\over 4\pi}\;=\;0.
\end{equation}
Using equation (\ref{L}), one computes
\begin{equation}\label{N}
r_h^2\left(Q^2 + P^2\right)\;=\;\left(r_h^2+a^2\right)\left[P^2+{16\pi^2\mu_B^2\left(r_h^2+a^2\right)L^2\over 9\Xi^2}-{8\pi\mu_BaPL\over 3\Xi}\right],
\end{equation}
and using this and equations (\ref{J}),(\ref{K}), and (\ref{LONDON}), one can show after some computation that (\ref{M}) can be expressed as
\begin{equation}\label{O}
{r_h^4\over L^2} -{2\pi \Xi r_h^3\over \varsigma_S}+\left[1+{4\pi\Gamma^2T^2L^2\over 9\Xi^2}\right]r_h^2+{L^2\over 4\pi \Xi^2}\left[BL^2-{4\pi\Gamma \, Ta\over 3}\right]^2 =0.
\end{equation}
This equation allows us to regard $r_h$ as a known function of the parameters describing the boundary field theory: in principle, given $a$, $\varsigma_S$, $\Gamma$, $T$, $B$ (and $L$), one solves this quartic for its largest root. (In practice, one does not regard $T$ as known: see below.)

Similar computations allow one to express equation (\ref{FROG}) in the following form:
\begin{eqnarray}\label{P}
{L^2\over 16\pi^2\Xi^2r_h^3}\left[{16\pi^2\Gamma^2T^2r_h^2\over 9}\,+\,\left(BL^2-{4\pi\Gamma \, Ta\over 3}\right)^2\right]\;+\;T \; \nonumber\\
\;-\; {r_h \Big(1\,+\,a^2/L^2\,+\,3r_h^2/L^2\,- \left(a^2/r_h^2\right)\Big)\over 4\pi (a^2\,+\,r_h^2)} \;=\; 0.
\end{eqnarray}
If we regard $a$, $\varsigma_S$, $\Gamma$, $B$, and $L$ as known, then equations (\ref{O}) and (\ref{P}) are a pair of simultaneous equations for $r_h$ and $T$ with known coefficients: so now $T$ (and $\mu_B\,=\,\Gamma\,T$) are, in principle, known, or at least computable, functions of these parameters. In particular, we can take a specific heavy-ion beam with a prescribed value of $\Gamma$ and of $\varsigma_S$, and consider the variation of $T$ and $\mu_B$ as $a$ and $B$ vary from collision to collision, allowing us to investigate the holography of the local rotation/magnetism analogy as it applies at various points of interest in the quark matter phase diagram.

It is evident that $T$ is not a simple function of these variables; in particular, it is an extremely complex function of $a$. In analysing this function, one begins by trying to understand $r_h$; but is not clear whether $r_h$ always increases or decreases with $a$ (other parameters being fixed) under these circumstances\footnote{In the special case where $\Gamma$ vanishes, it is not hard to show that $r_h$ does in fact always decrease as $a$ increases; but this is not clear if $\Gamma \neq 0.$}; and, even if that could be ascertained, it is still not clear whether $T$, given by equation (\ref{P}), increases or decreases with $a$. Bear in mind that we are \emph{not} fixing the black hole metric parameters (such as $M$) in this discussion, but rather the (somewhat distantly) related boundary parameters; so one cannot rely on familiar intuitions regarding the behaviour of the Hawking temperature under changes of black hole parameters. Similar comments apply to $\mu_B$.

In summary, then, holography reduces our questions regarding the local rotation/magnetism analogy to the solution of the pair of equations (\ref{O}) and (\ref{P}). These equations are relatively straightforward algebraic relations, so the simplification here is extreme. In practice, they are still sufficiently intricate as to require a numerical investigation, based on reasonably realistic data. We now proceed to that.

\addtocounter{section}{1}
\section* {\large{\textsf{3. Numerical Results}}}
For definiteness, we will focus on the region in the quark matter phase diagram around the proposed QGP/hadronic phase transition line, a major object of interest for various current and prospective beam energy scan experiments \cite{kn:shine,kn:nica,kn:fair,kn:BEAM,kn:luo}. We will try to apply holographic methods to data pertaining to locally rotating plasmas produced in such experiments. We stress again that the use of these data is merely to ensure that we are in the physical domain, not to try to make precise numerical predictions. The objective is to try to discern the \emph{trends}: do local rotation and magnetic fields tend to raise or lower the expected temperature and chemical potential (with all other parameters having fixed values)? Do they have these effects to similar degrees? Do they tend to change the shape of the phase line, and, again, if so, do they have effects of the same magnitude?
\addtocounter{section}{1}
\subsubsection* {\textsf{3.1. Rotation/Magnetism and the Critical Endpoint}}
We begin with a study of the effects of local rotation and magnetism on the location of the critical endpoint. The coordinates of this point are estimated in (for example) \cite{kn:karsch}. For definiteness we choose, as a rough estimate,
\begin{equation}\label{Q}
T_0^{CEP} \approx 140 \;\m{MeV}, \;\;\mu_{B\,0}^{CEP} \approx 280 \;\m{MeV},
\end{equation}
where the zero subscript indicates that these estimates effectively ignore the magnetic field (and therefore the angular momentum). This means that $\Gamma \approx 2$ in this Section.

We will investigate the way in which the endpoint is displaced by magnetic fields $B$ and  (specific) angular momenta $a$. Our principal objective at this stage is simply to assess which of the two effects is the dominant one. We stress that, in reality, $B$ and $a$ are by no means independent: they vary together in a complex manner as the impact parameter varies from collision to collision. For our purposes, however, it is more instructive to vary them independently. (An example in which a simplified version of their actual relationship is taken into account is provided in the next section.)

For the magnetic field, the classic study of Skokov et al. \cite{kn:skokov} finds that maximal values for the field in peripheral RHIC collisions are conveniently measured in units of the squared mass of the pion: let us call this $B_{\pi^2}$; note however that several subsequent studies have considered substantially larger values than those suggested in \cite{kn:skokov}. In order to gain some feel for the effect, and to make the trends clear, we have considered fields from zero (for central collisions) up to 15$B_{\pi^2}$; this value is probably realistic for collisions at the LHC, but probably \emph{not} for the experiments that probe the vicinity of the critical endpoint.

For the specific angular momentum, it is also difficult to estimate a realistic value: for a discussion of the reasons, see \cite{kn:86}. Maximal values around 75 fm are probably reasonable for RHIC collisions (considerably larger values may be possible at the LHC). We will settle on a conservative range between zero (again, corresponding to central collisions) and 50 fm.

With these assumptions, we can solve equations (\ref{O}) and (\ref{P}) numerically (for $r_h$ and $T$)\footnote{The value of $\varsigma_S$ is obtained by solving (for $\Gamma \approx 2$) the equations at $a = B = 0$. In solving the full versions of (\ref{O}) and (\ref{P}) one finds that there are in general two real solutions for $r_h$; one chooses the larger, since, as is well known, it is the outer horizon that determines the thermodynamics of the black hole. This identifies the correct solution for $T$.}. The results for the temperature (units MeV) are shown in the table, with values of $B$ increasing downward, and of $a$ increasing towards the right. (As above, we take it that the temperature is 140 MeV when $a = B = 0$.)
\begin{center}
\begin{tabular}{|c|c|c|c|c|c|c|}
  \hline
 & $a = 0 \;\;\m{fm}$   & 10 & 20 &30 & 40 & 50 \\
\hline
$B= 0 \times B_{\pi^2}$ &  140  & 137.9 & 131.5 & 121.1 & 106.5 & 88.3\\
$1$ &  139.9  & 137.9 & 131.5 & 121.0 & 106.5 & 88.2\\
$5$ &  139.2  & 137.0 & 130.5 & 119.7 & 104.4 & 84.2\\
$10$  &  136.7 & 134.4 & 127.3 & 115.3 &96.7 & 62.6\\
$15$  &  132.0  & 129.3  & 121.1 & 105.5 & 77.9 & < 40
 \\
\hline
\end{tabular}
\end{center}
We have expressed the result in the lower right corner as an inequality, on the grounds that in that (almost certainly unphysical) domain the numerical technique is not reliable.

Three aspects of the results are important:

$\bullet$ The effect of both rotational angular momentum and of magnetism is \emph{always} to reduce the temperature (and therefore, according to our model, the baryonic chemical potential) corresponding to the critical endpoint. This is not a trivial observation, because in the case of shearing angular momentum \cite{kn:86}, the effect of increasing $a$ (at non-zero $B$ and $\mu_B$) is sometimes to \emph{increase} the temperature somewhat\footnote{This answers the question raised in \cite{kn:86}: is this peculiar behaviour due to some aspect of the shearing motion itself? Evidently the answer is affirmative.} (though never above the value at $B = a = 0$.) Thus, the local rotation/magnetism analogy holds in this sense: both have the same kind of effect on the location of the critical endpoint. This effect is in agreement with previous (non-holographic) studies \cite{kn:magdy,kn:incera}.

$\bullet$ The effects are highly non-linear: for example, the drop in the temperature in going from $B= B_{\pi^2}$ to $B = 10 B_{\pi^2}$ is actually not as large as the drop in going from $B = 10 B_{\pi^2}$ to $B = 15 B_{\pi^2}$ (at any value of $a$), and similarly for the variation with $a$. This could lead to dramatic effects if the values of $B$ and $a$ prove to be larger in the beam energy scan experiments than one might expect.

$\bullet$ Most importantly, \emph{the effect of increasing $a$ completely dominates the effect of increasing} $B$: indeed, apart from the very extreme $B = 15 B_{\pi^2}$ case, the effects of increasing $B$ at $a = 0$ are almost certainly too small to be detectable, whereas rather modest values of $a$ at $B = 0$ lead to very appreciable drops in the expected temperature. To see this explicitly, compare Figure 2 (showing the effect on the temperature of increasing the magnetic field at a fixed value of $a$, $a = 30$ fm) with Figure 3 (showing the effect on the temperature of increasing $a$ at a fixed value of $B$, $B = 5 B_{\pi^2}$): the effect is obviously larger in the latter case\footnote{This is again in marked contrast to the case of shearing angular momentum, where similar values of $a$ lead to changes in the temperature of much the same size as those caused by increasing the magnetic field (both being very small).}.
\begin{figure}[!h]
\centering
\includegraphics[width=0.75\textwidth]{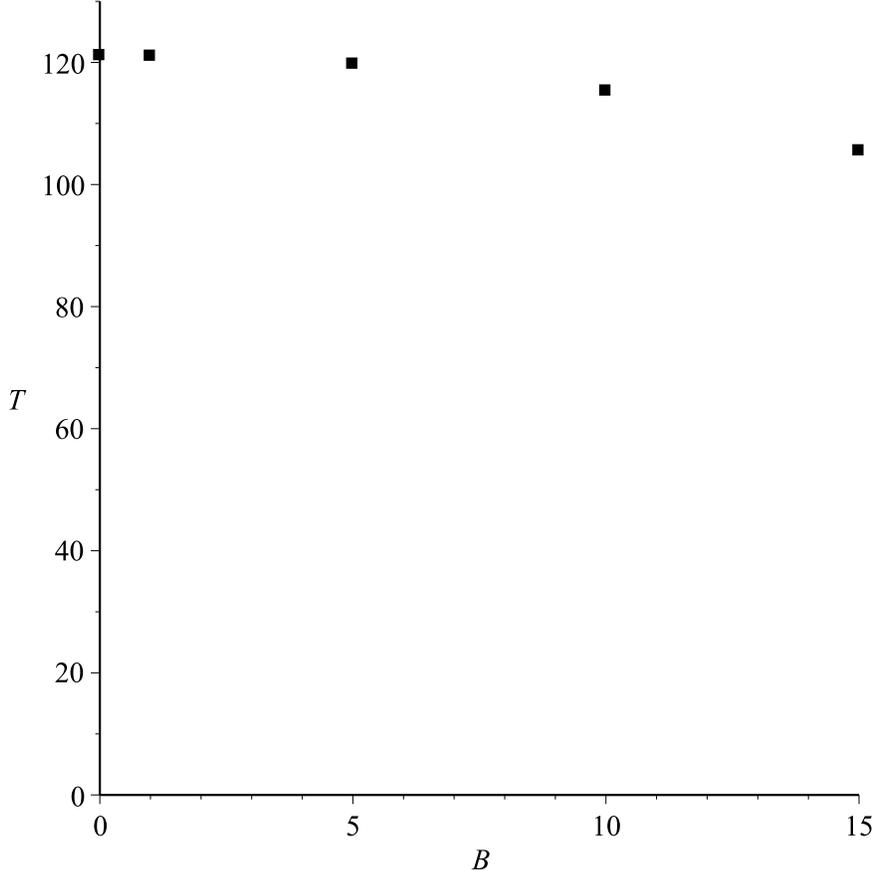}
\caption{Reduction of the temperature with increasing magnetic field $B$, at a fixed value of $a = 30$ fm. Units of $T$ are MeV, those of $B$ are $B_{\pi^2}$.}
\end{figure}
\begin{figure}[!h]
\centering
\includegraphics[width=0.75\textwidth]{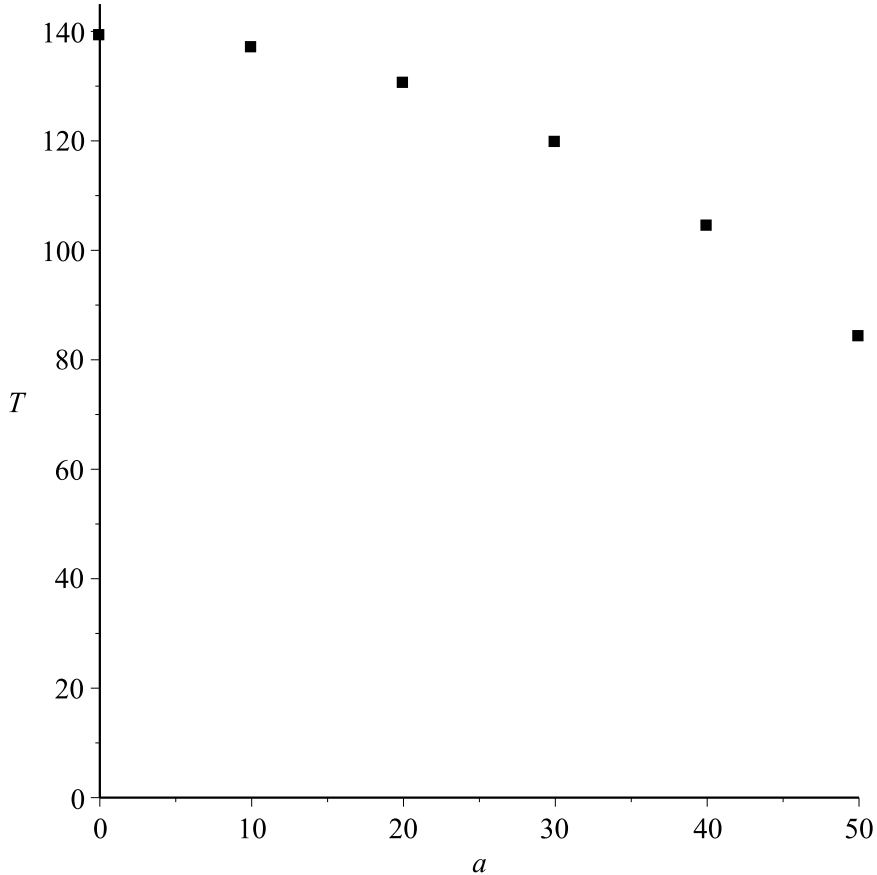}
\caption{Reduction of the temperature with increasing specific angular momentum $a$, at a fixed value of $B = 5 B_{\pi^2}$. Units of $T$ are MeV, those of $a$ are fm.}
\end{figure}

The qualitative conclusions here are twofold. First, holography indicates that the QGP produced in peripheral collisions in the beam energy scan experiments may have discernibly different properties to the plasma emerging from central collisions: in particular, the critical endpoint may be located, in the peripheral case, at unexpectedly low values of $T$ and $\mu_B$. Secondly, if such an effect should be observed, \emph{one will have to consider very carefully whether it is due to the magnetic field, and not rather to the vorticity in the plasma}. Ignoring this second possibility, clearly indicated by the gauge-gravity duality, could lead to severe over-estimates of the intensity of the magnetic field.

We conclude this section by noting that there has recently been much interest in the effect of \emph{finite system size} on computations of the location of the quark matter critical point; see for example \cite{kn:magdy} and numerous references therein. The point is that many earlier computations effectively assumed that the system was infinitely large: the question is whether taking finite size effects into account shifts the phase transition line towards or away from the origin of the phase diagram.

In the present case, a finite (and physically realistic) system size has been assumed from the outset ---$\,$ the vorticity is a \emph{local} rotational effect, and so we have applied our holographic model to small ``samples'' of the plasma. That is, the system size (related here to the parameters $a$ and $L$, both with dimensions of length) is already fixed by the basic physics of vorticity itself.

Nevertheless, one can ask whether holography throws any light on the question as to the way the phase line is shifted by varying the system size, that is, by varying $a$ and $L$. Care is required here: the systems we are dealing with involve plasmas moving locally (due to vorticity) at relativistic speeds, and the model must reflect this, while respecting causality. Concretely, our model has a dimensionless velocity parameter $a/L$, which must, of course, be smaller than unity (see the inequality (\ref{CATE})) but which should not be too small. Thus, if we wish to investigate the effects of increasing the size of the system, the natural way to do so is to increase $L$ while keeping $a/L$ fixed. That is, we should increase $L$, but also $a$ (and therefore $B$) appropriately.

If for example we consider a typical entry in the table above, say $a = 30$ fm, $B = 5 B_{\pi^2}$, and simply double $L$, $a$, and $B$, we find that the predicted temperature drops from 119.7 MeV to about 115.2 MeV. Similar effects are observed at all other values in the table. Thus, increasing the volume of the system causes, according to the holographic model, a contraction in the phase diagram towards the origin; broadly speaking, this is in agreement with the findings of \cite{kn:magdy} (and of some earlier investigations cited there). The effect is rather weak, however; interestingly, it varies with $a$ and $B$, being more pronounced at higher values of these parameters. It might be of interest to see whether this effect can be replicated with the methods of \cite{kn:magdy} (which focuses on the consequences of varying the system size at $B = 0$).

\addtocounter{section}{1}
\subsubsection* {\textsf{3.2. Rotation/Magnetism and the Shape of the Phase Line}}
We now wish to use the gauge-gravity duality to investigate the effect of local rotation/magnetism on the shape of the putative quark matter phase line. The curvature of this line has yet to be settled to the satisfaction of all, even in the immediate vicinity of the critical endpoint \cite{kn:karsch,kn:cea,kn:bonati1,kn:bonati2,kn:new,kn:newer}. We will therefore content ourselves here with the simplest possible question: do local rotation and magnetism tend to increase the magnitude of the curvature or decrease it?

We therefore begin with a sequence of five points in the phase diagram, arranged uniformly along a straight line within the region in the phase plane expected \cite{kn:fair,kn:de} to be explored at GSI-FAIR: roughly from the critical endpoint down to $T_0 \approx 100$ MeV, $\mu_{B\,0} \approx 600$ MeV, where, as before, zero subscripts indicate values taken at $a = B = 0$. Our objective is simply to see what becomes of this straight line.

We will abandon the fiction that the angular momentum and the magnetic field can be varied independently: in reality they vary together. We will not attempt to construct a realistic relation here; the simplest possible assumption, that the relation is linear, will suffice. In fact, the magnetic fields produced in the beam energy scan collisions are relatively small \cite{kn:incera}, but we will use somewhat unrealistically large values here, with the usual intention of clarifying the trend. (As we will see, the effect of magnetic fields here is in any case quite small relative to the effects of local rotation, so this does not unduly modify the results.) We consider, as in the preceding section, values of $a$ ranging from zero up to 50 fm, and corresponding values of $B$ ranging linearly with $a$ from zero up to $5 B_{\pi^2}$.

The results for the phase diagram coordinates of the five points, after they have been displaced by the effects of local rotation and magnetic fields, obtained as before by solving equations (\ref{O}) and (\ref{P}) with the appropriate parameter values, are given in the table, in the order $(T, \mu_B)$. For legibility we have deleted the units: they are MeV for $T$ and $\mu_B$, fm for $a$, and $B_{\pi^2}$ for $B$.
\begin{center}
\begin{tabular}{|c|c|c|c|c|c|}
  \hline
 & $a, B = 10,1$   & $a, B = 20, 2$ & $a, B = 30, 3$ &$a, B = 40, 4$& $a, B = 50, 5$ \\
\hline
$T_0, \mu_{B\,0} = 140, 280 $ &  137.9, 275.8   & 131.4, 262.8  & 120.58, 241.2  & 105.2, 210.4  & 84.2, 168.4  \\
$T_0, \mu_{B\,0} = 130, 360 $ &  127.8, 353.8  & 121.0, 335.2 & 109.9, 304.6 & 94.5, 261.8 & 74.4, 206.0 \\
$T_0, \mu_{B\,0} = 120, 440 $ &  117.7, 432.2 & 111.1, 407.6  & 100.2, 367.6  & 85.3, 313.2 & 66.7, 244.8 \\
$T_0, \mu_{B\,0} = 110, 520 $  &  107.8, 510.0 & 101.4, 479.8  & 91.1, 430.8  & 77.2, 365.2  &60.3, 285.2  \\
$T_0, \mu_{B\,0} = 100, 600 $  &  98.0, 587.8 & 92.0, 552.0   & 82.4, 494.4  & 69.7, 418.4  & 54.5, 327.2 \\
\hline
\end{tabular}
\end{center}
\begin{figure}[!h]
\centering
\includegraphics[width=0.75\textwidth]{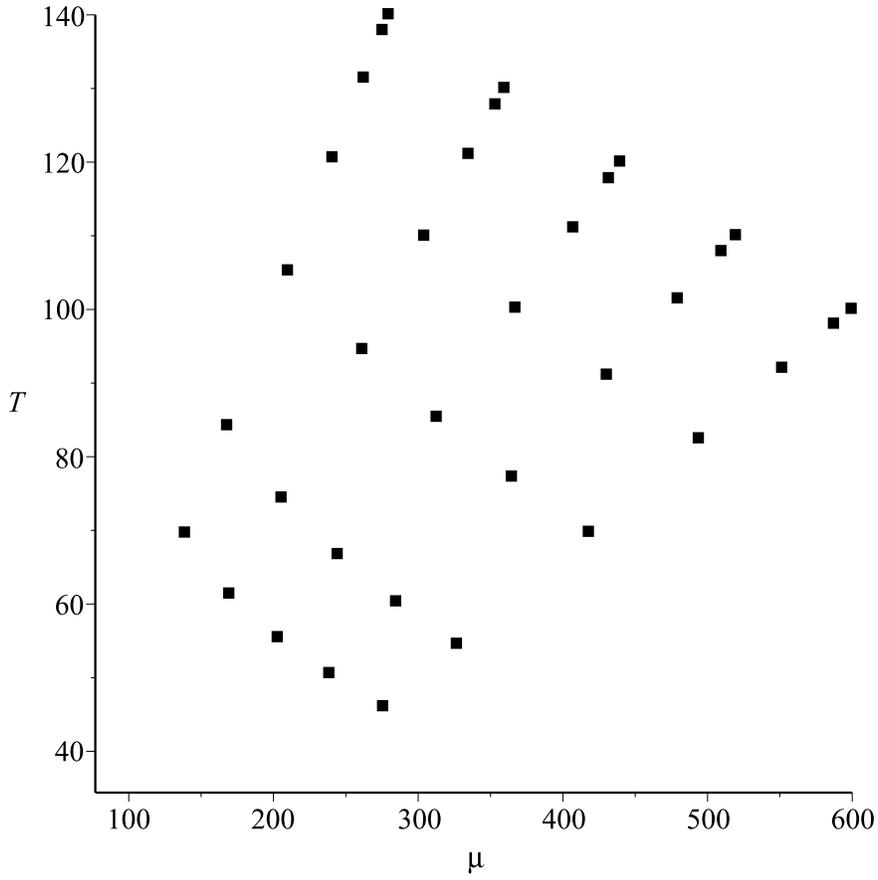}
\caption{Displacement and distortion of a line in the quark matter phase diagram by combined local rotation/magnetism. $(a, B) = (0,0)$ is the straight line at the top right; increasing values of $(a, B)$ displace this line towards the origin and cause its curvature to increase. The points at bottom left correspond to $a = 55$ fm, $B = 5.5 B_{\pi^2}$, for illustrative purposes (data not included in the Table). Units are MeV.}
\end{figure}
We need not go into a detailed analysis of these results (we have in fact carried out the calculations for many more points); instead we summarize as follows.

$\bullet$ For all values of $(a, B)$, the effect is to shift the line towards the origin; this is in agreement with the previous section and with general expectations \cite{kn:magdy,kn:incera}.

$\bullet$ Rotation/magnetism tends to cause the line to acquire a small amount of \emph{positive} curvature: while the line is pushed towards the origin, it bends upward (though never enough to compete with the downward translation). This effect increases with $(a, B)$.

$\bullet$ Not surprisingly in view of the results of the preceding section, a detailed analysis (not reflected in the Table) involving independent variation of $a$ and $B$ shows that, once again, angular momentum is the dominant effect here; magnetism is important only at completely unrealistic values of $B$.

This summary (with an additional line corresponding to $a = 55$ fm, $B = 5.5 B_{\pi^2}$, added for illustrative purposes) is represented in Figure 4.

In reality, of course, the phase line is not straight: it bends downward (see Figure 1, though note that the current estimates \cite{kn:karsch,kn:cea,kn:bonati1,kn:bonati2,kn:new,kn:newer} imply that the actual curvature is \emph{very much smaller} in magnitude than indicated there). We interpret our results as implying that the effect of local rotation/magnetism is to reduce the magnitude of this (negative) curvature.

In short, then, holography suggests that the QGP produced in some peripheral collisions may be described by a quark matter phase diagram which differs from the diagram for central collisions: the critical endpoint and its associated phase transition line may be displaced towards the origin, and the phase line may be appreciably straighter. (As we noted above, the deformation is small, but so is the magnitude of the initial curvature: for example, \cite{kn:new} and \cite{kn:newer} put it at around 0.01 or less, at least for values of $\mu_B$ which are not extremely large.) Once again, if such an effect were actually observed, one would have to find ways of determining whether it is due to the strong magnetic field or to the high vorticity associated with (some) such collisions. Holography points towards the latter.

\addtocounter{section}{1}
\section* {\large{\textsf{4. Conclusion}}}
In this work, we have used the gauge-gravity duality to investigate the combined effects of local rotation and strong magnetic fields on the QGP produced in some peripheral collisions. In reality, the internal motion of the plasma produced in such collisions is a complex mix of local rotation \emph{and} shearing, with one or the other dominating depending on physical conditions \cite{kn:KelvinHelm}.

The effects of shear in the QGP were considered, also from the gauge-gravity duality point of view, in \cite{kn:86}. They differ in many ways from those considered here. The principal difference is that, in the shearing case, the effects of angular momentum are quite small, generally even smaller than those of the magnetic field. Here, by contrast, rotational angular momenta have effects which can be much greater than those of the accompanying magnetic field. (On the other hand, they always have effects in the same direction, which, again, is not true of the shearing case.)

The question then arises: how can the effects of shear, vorticity, and magnetic fields be distinguished theoretically and in the data? In the gauge-gravity perspective, this is straightforward, since the various parameters enter the relevant equations in very different ways: for example, there is certainly no $a \leftrightarrow B$ symmetry in equations (\ref{O}) and (\ref{P}) above. Whether it will be easy to extract this distinction from the data, on the other hand, is open to doubt.

In overall summary, then: if effects like those we have found here (progressive shifting of critical endpoint and phase line towards the origin, and reduction of the magnitude of the curvature of the phase line) in peripheral collisions should actually be observed, then gauge-gravity duality indicates that they are probably due to (\emph{rotational}) angular momentum (vorticity) rather than to the magnetic field. If this is not borne in mind, there is a risk that the magnetic field will be over-estimated, with possible consequences for the study of the many phenomena expected to be associated with fields of that magnitude. One should perhaps be prepared to associate such observations instead with the chiral vortical effect \cite{kn:volosh} and allied phenomena.

\addtocounter{section}{1}
\section*{\large{\textsf{Acknowledgements}}}
The author wishes to acknowledge helpful discussions with Prof Soon Wanmei, and with Jude and Cate McInnes.

\addtocounter{section}{1}

\end{document}